\documentclass[12pt]{iopart}
\usepackage{graphicx, iopams,hyperref}
\begin{document}

\title[]{Entanglement in random pure states: Spectral density and average von Neumann entropy}

\author{Santosh Kumar and Akhilesh Pandey}

\address{School of Physical Sciences, Jawaharlal Nehru University, 
New Delhi - 110067, India}
\eads{\mailto{skumar.physics@gmail.com}, \mailto{ap0700@mail.jnu.ac.in}}
\begin{abstract}
Quantum entanglement plays a crucial role in quantum information, quantum teleportation and quantum computation. The information about the entanglement content between subsystems of the composite system is encoded in the Schmidt eigenvalues. We derive here closed expressions for the spectral density of Schmidt eigenvalues for all three invariant classes of random matrix ensembles. We also obtain exact results for average von Neumann entropy. We find that maximum average entanglement is achieved if the system belongs to the symplectic invariant class. 
\end{abstract}


\maketitle

%~~~~~~~~~~~~~~~~~~~~~ Section 1 ~~~~~~~~~~~~~~~~~~~~~~~~
%~~~~~~~~~~~~~~~~~~~~~~~~~~~~~~~~~~~~~~~~~~~~~~~~~~~~~~~~

\section{Introduction}

Quantum entanglement serves as the measure of quantum non-local correlations between the subsystems of a composite system. It has attracted a great deal of attention in recent times because of its crucial role in quantum information theory, quantum computation and quantum teleportation~\cite{HHHH1999,Reid2009}. Quantum entanglement has bizarre consequences which have baffled physicists over the years~\cite{EPR,Schro1935,Schro1936}. Despite this, quantum entanglement is now universally acknowledged as a useful resource which can be harnessed for a number of applications, e.g., superdense coding~\cite{BW,HHL}, quantum teleportation~\cite{Benn}, quantum cryptography~\cite{BB}, creation of a quantum computer~\cite{NC} etc.

In most cases it is desirable to have maximum possible entanglement between the subsystems of the compound system. It is known that random pure states lead to nearly maximal average entanglement. Most of the work so far has concentrated on random pure states which belong to the unitary invariant class ($\beta=2$)~\cite{Lub,Page,Sen,ZS,CSZ,Gir,KAT2008,ATK2009}, which corresponds to systems with broken time-reversal symmetry. It is natural to investigate time-reversal symmetric systems, i.e., random pure states belonging to the orthogonal ($\beta=1$) or symplectic ($\beta=4$) invariant classes. Recently there has been some work on systems with orthogonal invariance and now there are some important results available for systems with finite as well as large Hilbert space dimensions \cite{VMB,MBL,Vivo,Mjmdr2010}. For systems with symplectic invariance there has been work concerning only large dimensions and for extreme eigenvalue statistics \cite{VMB,CMV2010,CMV2011}. For these systems the distribution of Schmidt eigenvalues and various entanglement measures have not been investigated for finite dimesional cases.

Our purpose in this paper is to treat the random pure states belonging to all three invariant classes of random matrices~\cite{Mhta2004,For} on equal footing and obtain results valid for systems with arbitrary Hilbert space dimensions. We derive here the spectral density of Schmidt eigenvalues for arbitrary dimensions. We also derive the average von Neumann entropy and show that maximum average entanglement is achieved if one considers systems belonging to the symplectic invariant class. 

The presentation scheme in this paper is as follows. We begin with the brief introduction to the Wishart-Laguerre ensemble and give exact results for the spectral density in Section 2. In section 3 we briefly discuss the quantum entanglement problem in bipartite systems. Section 4 deals with the fixed-trace Wishart-Laguerre ensemble which is relevant for the entanglement problem. We obtain in this section closed results for the spectral density of Schmidt eigenvalues for all three invariant classes of random matrix ensembles. In section 5 we calculate the average von Neumann entropy. By comparing the von-Neumann entropies in the three cases, we show that, as far as average value is concerned, maximum entanglement is achieved for $\beta=4$. Finally, we conclude in section 6 with summary and some general remarks. Some proofs of the results are outlined in the appendices.

%~~~~~~~~~~~~~~~~~~~~~ Section 2 ~~~~~~~~~~~~~~~~~~~~~~~~
%~~~~~~~~~~~~~~~~~~~~~~~~~~~~~~~~~~~~~~~~~~~~~~~~~~~~~~~~~

\section{Wishart-Laguerre ensemble} 
\label{secFTLE}

Wishart-Laguerre ensembles were introduced by Wishart~\cite{Wsrt1928} in connection with the analysis of multivariate distributions. These ensembles comprise the $N$-dimensional matrices $\boldsymbol{H}=\boldsymbol{X}\boldsymbol{X}^{\dag}$ where the $\boldsymbol{X}$ are $N\times M$ dimensional matrices having its elements as independent and identical (iid) Gaussian random variables. 

Wishart-Laguerre ensembles arise explicitly in a number of problems in completely unrelated areas. Besides the quantum entanglement problem~\cite{ZS2001,BL2002,KTA2006} some other examples are chaotic mesoscopic systems~\cite{SN1994}, low energy QCD and gauge theories~\cite{Vrbst1994}, quantum gravity~\cite{AMK1994,Akmn1997}, quantitative finance~\cite{BP2001,BJ2004}, communication theory~\cite{Tltr1999,SKP2010a}, gene expression data analysis~\cite{ABB2000,HMMCBF2000} etc.

%~~~~~~~~~~~~~~~~~~ Subsection 2.1 ~~~~~~~~~~~~~~~~~~~~

\subsection {Joint probability density of eigenvalues}

Consider the elements of the matrices $\boldsymbol{X}$ constituting the Wishart-Laguerre ensemble $\boldsymbol{X}\boldsymbol{X}^{\dag}$ as real, complex or quaternion-real zero-mean iid Gaussian variables. These three cases lead to the three invariant classes of ensembles referred to as orthogonal, unitary and symplectic ensembles and are designated by the Dyson index $\beta=1, 2$ and 4~\cite{Mhta2004,For}. We choose $N\leq M$ for definiteness. The probability distribution followed by $\boldsymbol{X}$ is
\begin{equation}
\label{pd}
\boldsymbol{P}(\boldsymbol{X})\propto \exp\Big(-\frac{\boldsymbol{X}\boldsymbol{X}^{\dag}}{2v^2}\Big),
\end{equation}
where $v^2$ is the variance of each distinct real component of matrix elements of $\boldsymbol{X}$. The respective joint probability density (jpd) of eigenvalues ($x_j\in [0,\infty), j=1,...,N$) of $\boldsymbol{X}\boldsymbol{X}^{\dag}$ for the three cases come out as~\cite{For,Edlmn,SKP2009}
\begin{equation}
\label{jpd}
P^{(\beta)}(\{x\})=\mathrm{C}_{M,N}^{(\beta)}\left|\Delta_N(\{x\})\right|^\beta \exp\Big(-\sum_{i=1}^N x_i\Big)\prod_{j=1}^N x_j^\omega,
\end{equation}
where $\{x\}$ denotes $\{x_1,...,x_N\}$, $\Delta_N(\{x\})=\prod_{j<k}(x_j-x_k)$ is the Vandermonde determinant and the parameter $\omega$ is defined as 
\begin{equation}
\label{omg}
\omega=\frac{\beta}{2}(M-N+1)-1.
\end{equation}
The normalization $\mathrm{C}_{M,N}^{(\beta)}$ can be evaluated using the Selberg integral~\cite{Mhta2004} and turns out to be
\begin{equation}
\label{C}
\mathrm{C}_{M,N}^{(\beta)}=\prod_{j=1}^N\frac{\Gamma(\frac{\beta}{2}+1)}{\Gamma(\frac{\beta}{2}j+1)\Gamma(\frac{\beta}{2}(j-1)+\omega+1)}. 
\end{equation}
Note that (\ref{jpd}) is obtained from (\ref{pd}) for the choice $v^2=1/2$. Some authors choose $v^2=1$ (see for example~\cite{DE2002}), in which case apart from the change in normalization, the jpd in (\ref{jpd}) will have additional factor of 1/2 in the exponential.

%~~~~~~~~~~~~~~~~~~ Subsection 3.1 ~~~~~~~~~~~~~~~~~~~~

\subsection {Spectral densities for eigenvalues}
\label{sde}

In many cases one is interested in the study of quantities which are linear statistics on the eigenvalues. These quantities do not contain products of different eigenvalues. The averages of these can be evaluated using a single integral over the level-density of the eigenvalues, defined as
\begin{equation}
\label{R1}
 R_1^{(\beta)}(x)=N\int_0^\infty\cdots \int_0^\infty P^{(\beta)}(x,x_2,...,x_N)\,dx_2...dx_N.
\end{equation}
From random matrix theory~\cite{Mhta2004} we know that, for the classical ensembles, the above density can be obtained in terms of classical orthogonal polynomials for $\beta=2$ and in terms of classical skew-orthogonal polynomials for $\beta=1,4$~\cite{Dysn1972,NW,AFNM2000, PG2001,GP2002}.

For the Wishart-Laguerre ensemble defined by jpd~(\ref{jpd}), the exact expressions for level-densities are known for all $M,N$~\cite{SKP2011} in terms of associated Laguerre polynomials $L_\mu^{(b)}(x)$ ~\cite{Sz} and the corresponding weight function $w_b(x)=x^b e^{-x},~~ b>-1$.  
\noindent
Consider the parameter $a$ defined by 
\begin{equation}
\label{param}
2a+1=|M-N|.
\end{equation}
We have for unitary ensemble ($\beta=2$),
\begin{eqnarray}
\label{R1_ue}
\nonumber
\fl
R_1^{(2)}(N;a;x)=w_{2a+1}(x)\sum_{\mu=0}^{N-1}\frac{\Gamma(\mu+1)}{\Gamma(\mu+2a+2)}\Big(L_\mu^{(2a+1)}(x)\Big)^2\\
\!\!\!\!\!= w_{2a+1}(x) \frac{\Gamma(N+1)}{\Gamma(N+2a+1)}\left[L_{N-1}^{(2a+1)}(x)L_N^{(2a+2)}(x)-L_N^{(2a+1)}(x)L_{N-1}^{(2a+2)}(x)\right].
\end{eqnarray}
The second line in the above equation has been obtained using the Christoffel$-$Darboux sum~\cite{Sz}. Other expressions equivalent to the above sum are also possible; see (\ref{R1_ue_alt}). 

For orthogonal ensemble ($\beta=1$) the level density as well as higher order correlation functions are given in terms of skew-orthogonal polynomials~\cite{Dysn1972}. These skew-orthogonal polynomials can be expanded in terms of orthogonal polynomials~\cite{PG2001,GP2002}. Closed form expressions for the level density can be obtained using these expansions after some tedious algebra. We obtain for even $N$~\cite{SKP2011},
\begin{eqnarray}
\label{R1_oe_e}
\fl
\nonumber
R_{1,\mathrm{even}}^{(1)}(N;a;x)=2R_1^{(2)}(N;a;2x)\\
\nonumber
-w_{2a+1}(2x)L_{N-1}^{(2a+1)}(2x)\sum_{\mu=0}^{\frac{N-2}{2}}\frac{\Gamma\left(\frac{N+1}{2}\right)\Gamma(\mu+1)}{2^{2a}\Gamma\left(\frac{N+2a+1}{2}\right)\Gamma(\mu+a+2)}L_{2\mu+1}^{(2a+1)}(2x)\\ +w_a(x)L_{N-1}^{(2a+1)}(2x)\frac{\Gamma\left(\frac{N+1}{2}\right)}{\Gamma\left(\frac{N+2a+1}{2}\right)}\Big[\frac{2\Gamma(a+1,x)}{\Gamma(a+1)}-1\Big], ~~~~~
\end{eqnarray}
while for odd $N$ we get
\begin{eqnarray}
\label{R1_oe_o}
\fl
\nonumber
R_{1,\mathrm{odd}}^{(1)}(N;a;x)=2R_1^{(2)}(N;a;2x)\\
\nonumber
-w_{2a+1}(2x)L_{N-1}^{(2a+1)}(2x)\sum_{\mu=0}^{\frac{N-1}{2}}\frac{\Gamma\left(\frac{N+1}{2}\right)\Gamma(\mu+\frac{1}{2})}{2^{2a}\Gamma\left(\frac{N+2a+1}{2}\right)\Gamma(\mu+a+\frac{3}{2})}L_{2\mu}^{(2a+1)}(2x)\\ +w_a(x)L_{N-1}^{(2a+1)}(2x)\frac{\Gamma\left(\frac{N+1}{2}\right)}{\Gamma\left(\frac{N+2a+1}{2}\right)}, ~~~~~
\end{eqnarray}
The above two expressions may be put together as a single expression by defining 
\begin{equation}
c=N (\mbox{mod }2),
\end{equation}
so that we have for all $N$,
\begin{eqnarray}
\label{R1_oe}
\fl
\nonumber
R_1^{(1)}(N;a;x)=2R_1^{(2)}(N;a;2x)\\
\nonumber
-w_{2a+1}(2x)L_{N-1}^{(2a+1)}(2x)\sum_{\mu=0}^{\frac{N-2+c}{2}}\frac{\Gamma\left(\frac{N+1}{2}\right)\Gamma(\mu+1-\frac{c}{2})}{2^{2a}\Gamma\left(\frac{N+2a+1}{2}\right)\Gamma(\mu+a+2-\frac{c}{2})}L_{2\mu+1-c}^{(2a+1)}(2x)\\ +w_a(x)L_{N-1}^{(2a+1)}(2x)\frac{\Gamma\left(\frac{N+1}{2}\right)}{\Gamma\left(\frac{N+2a+1}{2}\right)}\big[(1-c)\frac{2\Gamma(a+1,x)}{\Gamma(a+1)}+(2c-1)\big]. ~~~~~
\end{eqnarray}
Here $\Gamma(b,x)$ represents the incomplete Gamma function, $\Gamma(b,x)=\int_x^\infty y^{b-1} e^{-y} dy$.

Finally we consider symplectic ensemble ($\beta=4$). Similar to $\beta=1$, the level density in this case also is given in terms of the corresponding skew-orthogonal polynomials. We use the expansion of these skew-orthogonal polynomials given in terms of orthogonal polynomials to obtain the following closed expression for the level-density~\cite{SKP2011}:
\begin{eqnarray}
\label{R1_se}
\nonumber
\fl
R_1^{(4)}(N;a;x)=\frac{1}{2}R_1^{(2)}(2N;2a+1/2;x)\\
-w_{4a+2}(x)L_{2N}^{(4a+2)}(x)\!\sum_{\mu=0}^{N-1}\frac{\Gamma(N+1)\Gamma(\mu+\frac{1}{2})}{2^{4a+3}\Gamma(N+2a+\frac{3}{2})\Gamma(\mu+2a+2)}L_{2\mu}^{(4a+2)}(x).
\end{eqnarray}
Note that in view of definition given by (\ref{param}), the parameter $a$ assumes only integral or half-integral values (the Wishart case), however, the above results hold for all real $a>-1$ (the Laguerre case).

Comparison between the above theoretical results and corresponding numerical simulations is shown in figure 1. They are in excellent agreement. 

\begin{figure*}[ht]
\centering
\includegraphics*[width=1.0 \textwidth]{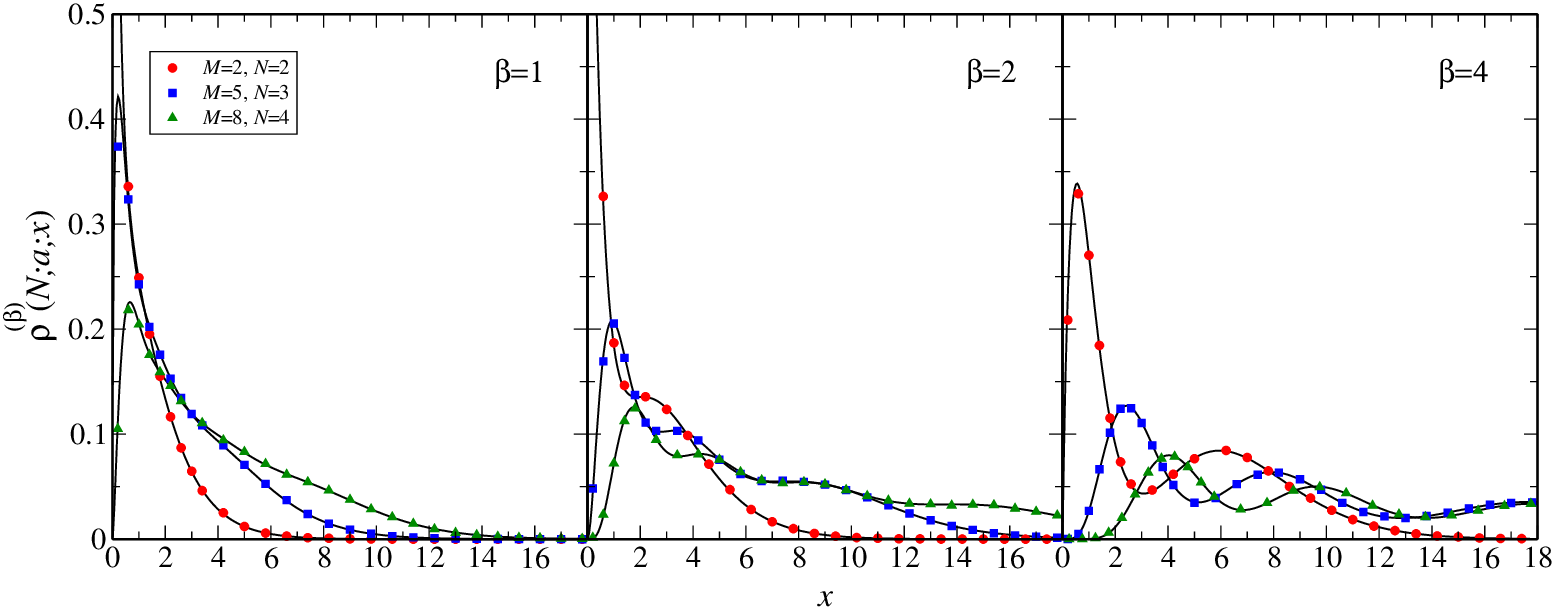}
\caption[]{Density, $\rho^{(\beta)}(N;a;x)=R_1^{(\beta)}(N;a;x)/N$, of eigenvalues for $\beta=1,2,4$. Solid lines are the theoretical predictions, whereas the symbols represent the simulation results.}
\end{figure*}

For large $N$, the density is described by the Mar\u{c}enko-Pastur formula~\cite{MP,VP},
\begin{equation}
\label{mp}
R_1^{(\beta)}(N;a;x)=\cases{
                      \frac{\sqrt{(x_{+}-x)(x-x_{-})}}{\pi\beta x}, & $x_{-}\leq x \leq x_{+}$,\\
		      ~~~~~~~0, & otherwise.                          
                     }
\end{equation}
Here $x_{\pm}=(\beta M/2)(1\pm\sqrt{N/M})^2$.

%~~~~~~~~~~~~~~~~~~~~~ Section 3 ~~~~~~~~~~~~~~~~~~~~~~~~
%~~~~~~~~~~~~~~~~~~~~~~~~~~~~~~~~~~~~~~~~~~~~~~~~~~~~~~~~~~

\section{Bipartite system}
\label{SecBS}

Consider a bipartite partition of an $NM$-dimensional Hilbert space $\mathcal{H}^{(NM)}$ consisting of subsystems $\mathbb{A}$ and $\mathbb{B}$ which span Hilbert spaces $\mathcal{H}^{(N)}_\mathbb{A}$ and $\mathcal{H}^{(M)}_\mathbb{B}$ such that $\mathcal{H}^{(NM)} = \mathcal{H}^{(N)}_\mathbb{A} \otimes \mathcal{H}^{(M)}_\mathbb{B}$. For
example, $\mathbb{A}$ may be a set of spins and $\mathbb{B}$ can be a heat bath. Without loss of generality we choose $M\ge N$. We take $|\Psi\rangle$ to be a quantum state of the composite system and $|i^\mathbb{A}\rangle$, $|\alpha^\mathbb{B}\rangle$ as the complete basis of $\mathcal{H}^{(N)}_\mathbb{A}$ and $\mathcal{H}^{(M)}_\mathbb{B}$ respectively. The state $|\Psi\rangle$ can then be expanded as the linear combination
\begin{equation}
|\Psi\rangle=\sum_{i=1}^N \sum_{\alpha=1}^M g_{i,\alpha}|i^\mathbb{A}\rangle \otimes |\alpha^\mathbb{B}\rangle.
\end{equation}
We focus on the situation where $|\Psi\rangle$ is a pure state, so that the density matrix of the composite system is given by $\rho=|\Psi\rangle \langle \Psi|$, with the constraint $\mbox{Tr}[\rho]=1$, which is equivalent to $\langle \Psi|\Psi\rangle=1$. 

The density matrix for the pure state $|\Psi\rangle$ can be written as
\begin{equation}
\label{dm}
\rho=\sum_{i=1}^N \sum_{\alpha=1}^M \sum_{j=1}^N \sum_{\gamma=1}^M g_{i,\alpha}g_{j,\gamma}^{*}|i^\mathbb{A}\rangle \langle j^\mathbb{A}|\otimes |\alpha^\mathbb{B}\rangle \langle\gamma^\mathbb{B}|
\end{equation}
The reduced density matrix for subsystem, say  $\mathbb{A}$, can be obtained by tracing out the subsystem $\mathbb{B}$. We obtain
\begin{equation}
\label{rdmA}
 \rho_\mathbb{A}=\sum_{i=1}^N \sum_{j=1}^N W_{i,j} |i^\mathbb{A}\rangle\langle j^\mathbb{A}|,
\end{equation}
where $W_{i,j}=\sum_{\alpha=1}^M g_{i,\alpha}g^{*}_{j,\alpha}$ can be viewed as the matrix elements of some matrix $\boldsymbol{W}=\boldsymbol{G}\boldsymbol{G}^\dag$, $\boldsymbol{G}$ being $N\times M$ dimensional. In the diagonal basis, (\ref{rdmA}) can be written as
\begin{equation}
\rho_\mathbb{A}=\sum_{i=1}^N \lambda_i \,|\lambda_i^\mathbb{A}\rangle\langle \lambda_i^\mathbb{A}|.
\end{equation}
The $\lambda_i$ are referred to as Schmidt eigenvalues. The trace condition restricts the Schmidt eigenvalues to the following constraint
\begin{equation}
\label{ftc}
 \sum_{i=1}^N \lambda_i=\mathrm{tr} (\boldsymbol{G}\boldsymbol{G}^\dag)=1.
\end{equation}
$|\Psi\rangle$ is called a random pure state if the coefficients $g_{i,\alpha}$ are chosen at random from some given probability distribution.
 
 %Note that the matrices $G$ obey $\boldsymbol{\mathcal{P}}(\boldsymbol{G})\propto \delta(\mathrm{Tr}(\boldsymbol{G}\boldsymbol{G}^\dag)-1)$.

%~~~~~~~~~~~~~~~~~~~~~ Section 4 ~~~~~~~~~~~~~~~~~~~~~~~~
%~~~~~~~~~~~~~~~~~~~~~~~~~~~~~~~~~~~~~~~~~~~~~~~~~~~~~~~~~
\section{Fixed trace Wishart-Laguerre ensembles}

As discussed in the previous section, the Schmidt eigenvalues correspond to Wishart type of matrices. However, the conservation of probabilities for quantum states constrains the trace of these matrices. In earlier works Wishart-Laguerre ensembles have been used to model random quantum states~\cite{BL2002,KTA2006}. However, they do not provide a complete quantitative description of the random quantum states due to the neglect of fixed trace condition~\cite{KTA2006}. The deviations are prominent particularly when the Hilbert space-dimension of the smaller subsystem is not large. Detailed discussion of superiority of fixed-trace ensemble over the ordinary Wishart-Laguerre ensemble for describing random quantum states can be found in~\cite{KAT2008,ATK2009}.

%~~~~~~~~~~~~~~~~~~ Subsection 4.1 ~~~~~~~~~~~~~~~~~~~~
\subsection {Joint probability density of Schmidt eigenvalues}

Consider the matrix elements $g_{i,\alpha}$ as iid (real, complex or quaternion-real) Gaussian variables with zero mean, such that the fixed trace condition (\ref{ftc}) is satisfied. The probability distribution for $\boldsymbol{G}$ is given by
\begin{equation}
\boldsymbol{\mathcal{P}}(\boldsymbol{G})\propto \delta(\mathrm{tr}(\boldsymbol{G}\boldsymbol{G}^\dag)-1). 
\end{equation}
Correspondingly the jpd of Schmidt eigenvalues ($\lambda_j\in [0,1], j=1,...,N$) is obtained as
\begin{equation}
\label{ftjpd}
 \mathcal{P}^{(\beta)}(\{\lambda\})=\mathcal{C}_{M,N}^{(\beta)} \delta\Big(\sum_{i=1}^N \lambda_i-1\Big)\left|\Delta_N(\{\lambda\})\right|^\beta \prod_{j=1}^N \lambda_j^\omega.
\end{equation}
Note that, with the fixed trace condition, (\ref{ftc}), the exponential term in (\ref{jpd}) becomes a constant and one is left with (\ref{ftjpd}).
The normalization constant in the above equation is
\begin{equation}
\label{Cft}
 \mathcal{C}_{M,N}^{(\beta)}=\Gamma\Big(\frac{\beta M N}{2}\Big) \mathrm{C}_{M,N}^{(\beta)}~,
\end{equation}
where $\mathrm{C}_{M,N}^{(\beta)}$ is given by (\ref{C}); see \ref{CCft}. The three $\beta$ values (1, 2, 4) in (\ref{ftjpd}) again correspond to the cases when the coefficients $g_{i,\alpha}$ in (\ref{dm}) are real, complex or quaternion-real. Also, the choice of variance of Gaussian elements is immaterial in this case.

%~~~~~~~~~~~~~~~~~~ Subsection 4.1 ~~~~~~~~~~~~~~~~~~~~

\subsection {Spectral densities for Schmidt eigenvalues}
\label{secSDSE}

The level-density of Schmidt eigenvalues is,
\begin{equation}
\label{R1ft}
\mathcal{R}_1^{(\beta)}(M,N;\lambda)=N\int_0^\infty\cdots \int_0^\infty \mathcal{P}^{(\beta)}(\lambda,\lambda_2,...,\lambda_N)\,d\lambda_2...d\lambda_N.
\end{equation}
Note that because of the presence of delta function constraint the original limits of integration $[0,1]$ can be replaced by $[0,\infty)$.

We give here the exact expressions for the level density of Schmidt eigenvalues obtained from the jpd~(\ref{ftjpd}). Proof is briefly outlined in Appendices B and C.

We need the following to express the desired results compactly:
 
\begin{equation}
\fl
A_\mu^{\xi,m,n}=\frac{(-1)^\mu \Gamma(\frac{\xi}{2} m+1)\Gamma(\frac{\xi}{2}m n)}{\frac{\xi}{2}\Gamma(\mu+1)\Gamma(\frac{\xi}{2}n-\mu)\Gamma(\frac{\xi}{2}(mn-m+n)-\mu-1)\Gamma(\mu+\frac{\xi}{2}(m-n)+2)},
\end{equation}
\begin{eqnarray}
\fl
B_{\mu}^{\,m,n}=\frac{(-1)^\mu 2^{n+1}\Gamma(\frac{m n}{2})\Gamma(\frac{m+1}{2})\Gamma(\frac{n+1}{2})}{\pi^{1/2}\Gamma(\frac{m-n+1}{2})\Gamma(\mu+m-n+1)\Gamma(\mu+1)\Gamma(n-\mu)},
\end{eqnarray}
\begin{eqnarray}
\fl 
C_{\mu,\nu}^{m,n}=\frac{2^{m-n-1}\Gamma(\frac{m-n+1}{2})\Gamma(\nu+\frac{m-n+2-c}{2})}{\Gamma(\frac{m n}{2}-m+n-\mu-1)\Gamma(\nu+\frac{3-c}{2})}B_{\mu}^{\,m,n},
\end{eqnarray}
\begin{equation}
\fl
D^{m,n}=\frac{2^{m-n}\Gamma(\frac{m+1}{2})\Gamma(\frac{m n}{2})}{\Gamma(\frac{(m+1)(n-1)}{2})\Gamma(\frac{n}{2})},
\end{equation}
\begin{eqnarray}
\fl
K_{\mu,\nu}^{m,n}=\frac{2^{2\mu-2N}(\nu+2m-2n+1)\Gamma(2n-\nu)\Gamma(\mu+\frac{1}{2})\Gamma(\mu+m-n+\frac{1}{2})}{\Gamma(m+\frac{1}{2})\Gamma(n+\frac{1}{2})\Gamma(2\mu-\nu+1)}A_{\nu}^{4,m,n},
\end{eqnarray}
\begin{eqnarray}
\label{Smu}
\nonumber
\fl
\mathcal{S}_\mu(\xi,\eta,b;m,n;y)=y ^{\,\mu +\frac{\xi }{2}(m-n)}(1-y)^{-\mu+\frac{\xi}{2}(m n-m+n)-2 }\\
\times~_2\widetilde{F}_1\Big(\eta -\frac{\xi }{2} n,\mu -\frac{\xi}{2}(m+1)(n-1);1+\frac{\xi }{2}(m-n);\frac{b\,y }{y-1}\Big),
\end{eqnarray}
\begin{equation}
 \fl 
 \mathcal{T}_\mu(\xi;m,n;y)=A_\mu^{\xi,m,n}\Big[\frac{\xi}{2}n\, \mathcal{S}(\mu,\xi,1,1;m,n,y)-\Big(\frac{\xi}{2}n-\mu-1\Big) \mathcal{S}(\mu,\xi,0,1;m,n;y)\Big],~
\end{equation}
\begin{eqnarray}
\nonumber
\fl
\mathcal{U}_\mu(m,n;y)=(2y) ^{\,\mu+m-n+1}(1-2y)^{-\mu+\frac{mn}{2}-m+n-1}\\
\times~_2\widetilde{F}_1\Big(1,-\frac{1}{2}(m-n-1);\frac{mn}{2}-m+n-\mu;2-\frac{1}{y}\Big).
\end{eqnarray}
Here $~_2\widetilde{F}_1(l,m;n;y)=~_2F_1(l,m;n;y)/\Gamma(c)$ is the regularized hypergeometric function.

We first consider the unitary case ($\beta=2$), which has been considered by several authors. In~\cite{KAT2008} the result was announced for the square case ($M=N$) in terms of sum involving hypergeometric$~_3F_2$. Later the result was extended to rectangular case and was given in terms of sum comprising hypergeometric$~_5F_4$~\cite{ATK2009}. In~\cite{Vivo} a triple sum expression has been given. We give here a result, valid for all $M\geq N$, which is much simpler than those obtained earlier and is given as a single sum over the hypergeometric$~_2F_1$. We have
\begin{eqnarray}
\label{R1ft2}
\fl
\nonumber
\mathcal{R}_1^{(2)}(M,N;\lambda)=\sum_{\mu=0}^{N-1}  \mathcal{T}_\mu(2;M,N;\lambda),\\
\end{eqnarray}

The spectral density for the orthogonal case ($\beta=1$) has been obtained for $N$ even in~\cite{Vivo}. We consider here both even and odd $N$ cases. For even $N$, we obtain
\begin{eqnarray}
\label{R1ft1e}
\fl
\nonumber
\mathcal{R}_{1,\mathrm{even}}^{(1)}(M,N;\lambda) =\Big[4\sum_{\mu=0}^{N-1}\mathcal{T}_\mu\Big(4;\frac{M}{2},\frac{N}{2};2\lambda\Big)
+\sum_{\mu=0}^{N-1} B_{\mu}^{M,N}\mathcal{U}_\mu(M,N,\lambda)\\
\nonumber
-\sum_{\mu=0}^{N-1}\sum_{\nu=0}^{\frac{N-2}{2}}C_{\mu,\nu}^{M,N}\mathcal{S}_\mu\Big(4,N-2\nu-1,1;\frac{M}{2},\frac{N}{2};2\lambda\Big)\Big]\Theta\left(\frac{1}{2}-\lambda\right)\\
-D^{M,N}\mathcal{S}_{\frac{-M+N-1}{2}}\Big(4,1,2;\frac{M}{2},\frac{N}{2};\lambda\Big),
\end{eqnarray}
while for odd $N$ we have,
\begin{eqnarray}
\label{R1ft1o}
\fl
\nonumber
\mathcal{R}_{1,\mathrm{odd}}^{(1)}(M,N;\lambda) =\Big[4\sum_{\mu=0}^{N-1}\mathcal{T}_\mu\Big(4;\frac{M}{2},\frac{N}{2};2\lambda\Big)\\
\nonumber
-\sum_{\mu=0}^{N-1}\sum_{\nu=0}^{\frac{N-1}{2}}C_{\mu,\nu}^{M,N}\mathcal{S}_\mu\Big(4,N-2\nu,1;\frac{M}{2},\frac{N}{2};2\lambda\Big)\Big]\Theta\left(\frac{1}{2}-\lambda\right)\\
+D^{M,N}\mathcal{S}_{\frac{-M+N-1}{2}}\Big(4,1,2;\frac{M}{2},\frac{N}{2};\lambda\Big).
\end{eqnarray}
Similar to (\ref{R1_oe}) the two expressions for even and odd cases may be combined into one as,
\begin{eqnarray}
\label{R1ft1}
\fl
\nonumber
\mathcal{R}_1^{(1)}(M,N;\lambda) =\Big[4\sum_{\mu=0}^{N-1}\mathcal{T}_\mu\Big(4;\frac{M}{2},\frac{N}{2};2\lambda\Big)
+(1-c)\sum_{\mu=0}^{N-1} B_{\mu}^{M,N}\mathcal{U}_\mu(M,N,\lambda)\\
\nonumber
-\sum_{\mu=0}^{N-1}\sum_{\nu=0}^{\frac{N-2+c}{2}}C_{\mu,\nu}^{M,N}\mathcal{S}_\mu\Big(4,N-2\nu-1+c,1;\frac{M}{2},\frac{N}{2};2\lambda\Big)\Big]\Theta\left(\frac{1}{2}-\lambda\right)\\
-(-1)^{N}D^{M,N}\mathcal{S}_{\frac{-M+N-1}{2}}\Big(4,1,2;\frac{M}{2},\frac{N}{2};\lambda\Big),
\end{eqnarray}
$\Theta(y)$ in the above equations is the Heaviside theta function, being 0, 1 respectively for $y<0$ and $y>0$.

Finally, we consider the symplectic case ($\beta=4$). Note that we do not take into account the Kramers degeneracy explicitly and use the jpd given by ~(\ref{ftjpd}). We obtain
\begin{eqnarray}
\label{R1ft4}
\fl
\mathcal{R}_1^{(4)}(M,N;\lambda)=\sum_{\mu=0}^{2N-1}\mathcal{T}(4;M,N;\lambda)-\sum_{\mu=0}^{N-1}\sum_{\nu=0}^{2\mu} K_{\mu,\nu}^{M,N}\mathcal{S}_\nu(4,0,1;M,N;\lambda).
\end{eqnarray}

\begin{figure*}[ht]
\centering
\includegraphics*[width=0.9 \textwidth]{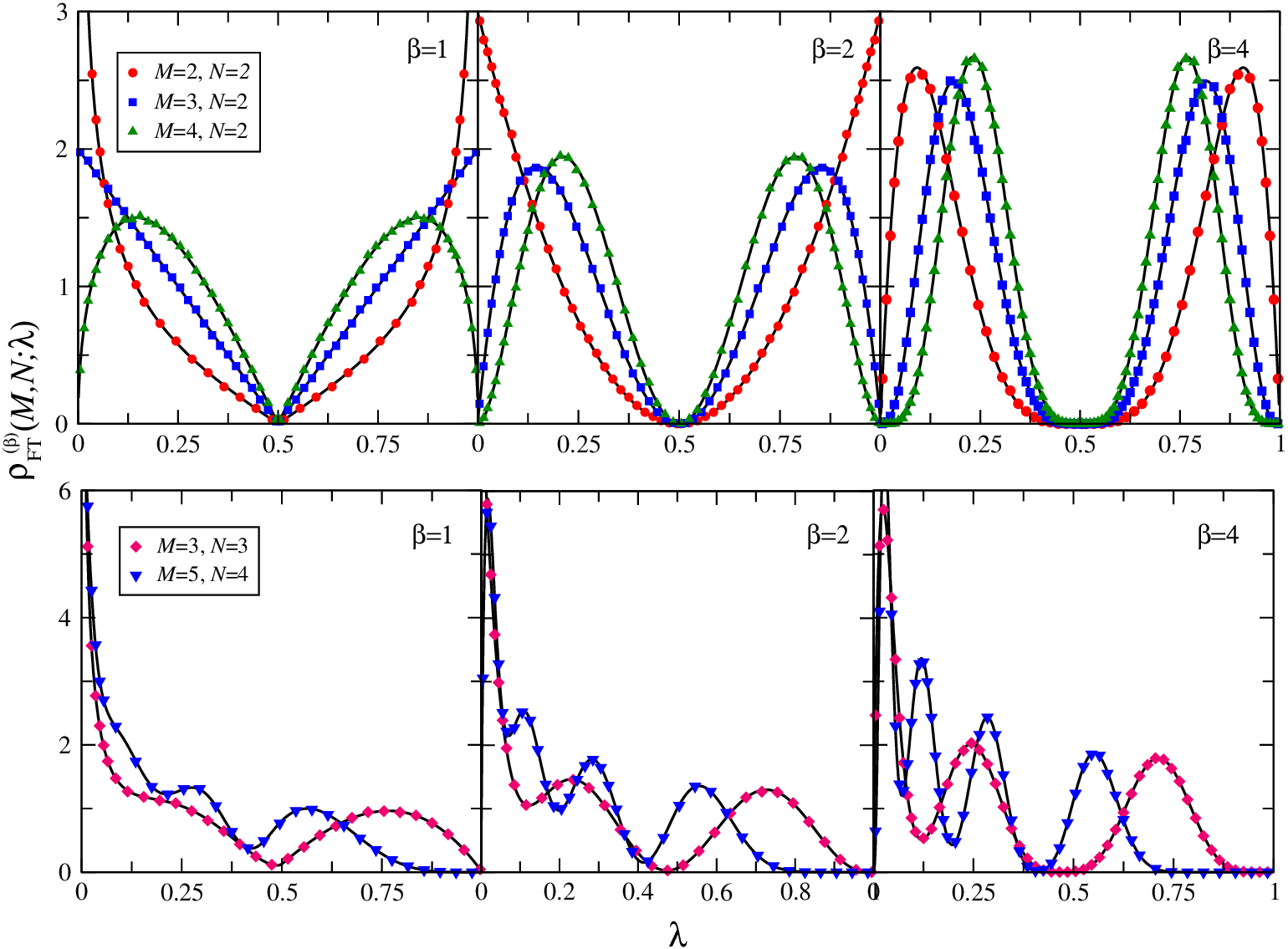}
\caption[]{Density, $\rho_{\mathrm{FT}}^{(\beta)}(M,N;\lambda)=\mathcal{R}_1^{(\beta)}(M,N;\lambda)/N$, of Schmidt eigenvalues for $\beta=1,2,4$ for several $M, N$ values. Solid lines are the theoretical predictions, whereas the symbols represent the simulation results. }
\label{fig-schmit}
\end{figure*}

Of particular interest is the case of $N=2$ and arbitrary $M\ge 2$, as it corresponds to a physical situation where a single qubit is entangled to a heat bath. In this case the expression for level density is simple and can be obtained directly from (\ref{ftjpd}). We have for all three $\beta$,
\begin{equation}
\label{densN2}
\mathcal{R}_1^{(\beta)}(M,2;\lambda)=\kappa^{(\beta,M)} \lambda^{\frac{\beta}{2}(M-1)-1}(1-\lambda)^{\frac{\beta}{2}(M-1)-1}|1-2\lambda|^\beta,
\end{equation}
where $\kappa^{(\beta,M)}$ is given by
\begin{equation}
\kappa^{(\beta,M)}=\cases{
             2^{M-2}(M-1), & $\beta=1$,\\
	     \frac{(M-1)\Gamma(2M)}{(\Gamma(M))^2}, & $\beta=2$,\\
	     \frac{(2M-1)(2M-2)\Gamma(4M)}{6(\Gamma(2M))^2}, & $\beta=4$.
            }
\end{equation}
For large $M$, $\lambda$ close to 1/2 dominates the level density (\ref{densN2}) and we obtain,
\begin{equation}
 \mathcal{R}_1^{(\beta)}(M,2;\lambda)\approx \frac{(\beta M)^{\frac{\beta+1}{2}}}{2^{\frac{\beta-3}{2}}\Gamma(\frac{\beta+1}{2})}\, e^{-\frac{\beta M}{2}(1-2\lambda)^2} |1-2\lambda|^\beta.
\end{equation}
As can be seen in figure \ref{fig-schmit} as $M$ increases the two maxima move closer and closer towards $\lambda=1/2$ and become sharply peaked like delta function for $M\rightarrow \infty$.
 
We have numerically generated fixed trace Wishart-Laguerre ensemble by considering matrices $\boldsymbol{X}\boldsymbol{X}^\dag/\mathrm{tr}(\boldsymbol{X}\boldsymbol{X}^\dag)$, where $\boldsymbol{X}\boldsymbol{X}^\dag$ as above form the Wishart-Laguerre ensemble. Comparison of the above theoretical results with numerics is shown in figure. They agree perfectly.

The level densities obtained above give good description of the Schmidt eigenvalues. These expressions are important because the statistical properties of entanglement are encoded in the Schmidt eigenvalues. For the unitary case the one-level density has been found to describe accurately the distribution of the Schmidt eigenvalues over the whole range for coupled quantum systems exhibiting chaos \cite{KAT2008}. This confirms the existence of universality for the distribution of the
Schmidt eigenvalues in quantum chaos. The level-densities can be used to calculate the averages of quantities which are linear statistics on Schmidt eigenvalues, e.g., purity, von Neumann entropy, R\'{e}nyi entropy etc., thereby giving insight into the extent of entanglement.

%~~~~~~~~~~~~~~~~~~~~~ Section 5 ~~~~~~~~~~~~~~~~~~~~~~~~

\section {Average von Neumann Entropy}
\label{secAVNE}

As discussed above, the information about the degree of entanglement is stored in the Schmidt eigenvalues. One can construct a suitable measure of entanglement by considering a function of these eigenvalues whose value will tell us how entangled a pure state is. Von Neumann entropy serves as an important measure of degree of entanglement. It is an extension of classical entropy concepts to the field of quantum mechanics. In terms of the Schmidt eigenvalues, it is defined as 
\begin{equation}
\label{VNE}
\mathcal{E}^{(\beta)}=-\sum_{j=1}^N \lambda_j \ln \lambda_j. 
\end{equation}
It acquires the maximum value of $\ln N$, when each of the Scmidt eigenvalues assume the value $1/N$. On the other hand the minimum value of 0 is obtained when one of the eigenvalues is one and rest are zero. These two scenarios correspond respectively to maximally entangled and completely unentangled (separable) cases.

To calculate the average von Neumann entropy, $E^{(\beta)}=\langle \mathcal{E}^{(\beta)} \rangle$, we need to perform the following integral:
\begin{equation}
\label{E1beta}
E^{(\beta)}(M,N)=-\int_0^\infty \cdots \int_0^\infty\Big(\sum_{i=1}^N \lambda_i \ln \lambda_i\Big)\mathcal{P}^{(\beta)}(\{\lambda\})\,d\lambda_1...d\lambda_N.~~~~~
\end{equation}
As in (\ref{R1ft}), we have again considered the limits of integrations as $[0,\infty)$ instead of $[0,1]$. Von Neumann entropy being a linear statistic, the symmetry of the eigenvalues in jpd allows us to reduce the above average involving $N$ integrals to an average involving a single integral, i.e.,
\begin{equation}
E^{(\beta)}(M,N)=-\int_0^\infty\lambda \ln \lambda \;\mathcal{R}^{(\beta)}(M,N;\lambda)\,d\lambda.
\end{equation}
The above calculation can be done using the results for level density given in preceding section. However, it turns out that the above average over the fixed trace ensemble can be reduced directly to an average over the ordinary Wishart-Laguerre ensemble by introducing an auxiliary gamma function integral~\cite{Page}. The latter route to calculation of average von-Neumann entropy is somewhat more straightforward and we follow it below.

As shown in \ref{AGFIM} we have,
\begin{equation}
\label{E2beta}
E^{(\beta)}(M,N)=
\psi\Big(\frac{\beta M N}{2}+1\Big)-\frac{2}{\beta M N}\int_0^\infty\!\!\!x \ln x\, R_1^{(\beta)}(N;a;x)\,dx.
\end{equation}
where $R_1^{(\beta)}(N;a;x)$ is the level density for ordinary Wishart-Laguerre ensemble, as given in (\ref{R1}) and $\psi(x)$ represents the digamma function defined by
\begin{equation}
\label{digamma}
\psi(y)=\frac{1}{\Gamma(y)}\int_0^\infty e^{-r} r^{y-1} \ln r\,dr.
\end{equation}
Equation (\ref{E2beta}) gives a compact expression for the average von Neumann entropy for all three $\beta$ values. For $M\geq N>>1$, the average von Neumann entropy can be found from (\ref{E2beta}) using the Mar\u{c}enko-Pastur density given by (\ref{mp}) . Also in this limit 
\begin{equation}
\label{psi_asym}
\psi\Big(\frac{\beta M N}{2}\Big)=\ln\Big(\frac{\beta M N}{2}\Big)+\mathcal{O}\Big(\frac{1}{MN}\Big). 
\end{equation}
Using (\ref{mp}), the integral in (\ref{E2beta}) is obtained as $(\beta M N/2)\ln(\beta M/2)+\beta N^2/4$, thereby giving the large $N$ result for all three $\beta$ as
\begin{equation}
\label{Ebeta_asym}
E^{(\beta)}(M,N)=\ln(N)-\frac{N}{2M}. 
\end{equation}
The above result was obtained for $\beta=2$ in \cite{Page}. See \cite{BL2002} for $\beta=1$.

 \begin{figure*}[ht]
\centering
\includegraphics*[width=0.9 \textwidth]{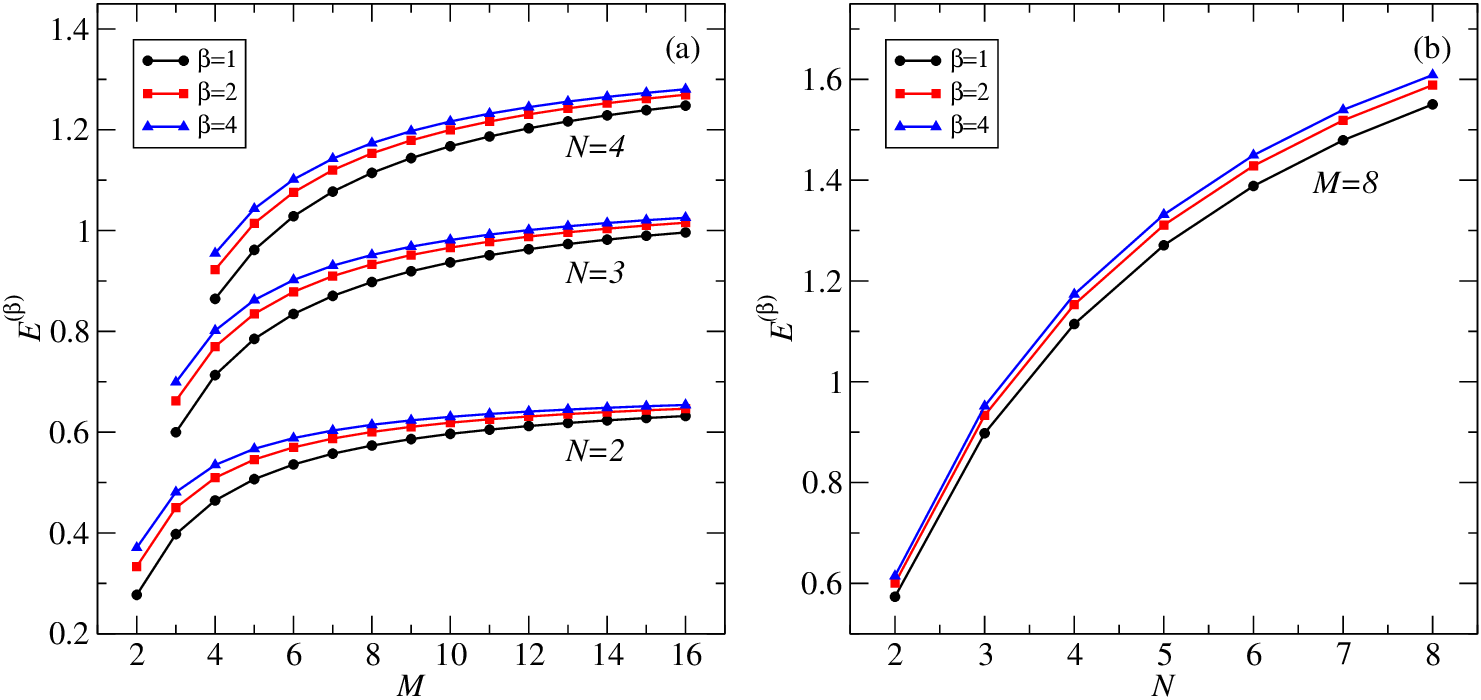}
\caption[]{Average von Neumann entropy for (a) $N=2, 3, 4$ and $M=2$ to 16 (b) $M=8$ and $N=2$ to 8.}
\label{avne}
\end{figure*}

Exact closed form expressions for the average von Neumann entropy for arbitrary $M,N$ can be calculated using (\ref{E2beta}) and the results in section \ref{sde}. To evaluate the integral in (\ref{E2beta}) we use the following identity:
\begin{equation}
\int_{0}^{\infty}x^b \ln(x) f(b,x)\,dx=\frac{\partial}{\partial b}\int_{0}^{\infty} x^b f(b,x)\,dx-\int_{0}^{\infty} x^b \frac{\partial f(b,x)}{\partial b}\,dx.
\end{equation}
The evaluation of the above integral for the three invariant classes involves the use of properties followed by Laguerre polynomials \cite{Sz} and its expansion (\ref{Laguerre}). 

The average von Neumann entropy in $\beta=2$ case is already known \cite{Page,Sen}. We have,
\begin{eqnarray}
\label{ent2}
E^{(2)}(M,N)=\sum_{\mu=M+1}^{M N}\frac{1}{\mu}-\frac{N-1}{2M}. 
\end{eqnarray}
For $\beta=1$, even $N$, we obtain
\begin{eqnarray}
\label{ent1e}
\nonumber
\fl
&&E^{(1)}=\psi\Big(\frac{M N}{2}+1\Big)-\psi(M)-\frac{(N-1)}{2M}+\frac{(M-1)(N-1)}{M N}\ln2+\frac{2^{N-1}\Gamma(\frac{M+1}{2})\Gamma(\frac{N+1}{2})}{M N\sqrt{\pi}}\\
\nonumber
\fl
&&\times\sum_{\mu=0}^{\frac{N-2}{2}}\sum_{\nu=0}^{N-1}\frac{(-1)^\nu (\nu+M-N+1)\Gamma(\mu+1)\psi(\nu+M-N+1)(-\nu-1)_{2\mu+1}}{\Gamma(\mu+\frac{M-N+3}{2})\Gamma(N-\nu)\Gamma(\nu+1)\Gamma(2\mu+2)}\\
\nonumber
\fl
&&-\frac{2^{\frac{M+N+d}{2}}\Gamma(\frac{M+1}{2})\Gamma(\frac{N+1}{2})}{M N\sqrt{\pi}}\\
\nonumber
\fl
&&\times\sum_{\mu=1}^{\frac{M-N+d}{2}}\sum_{\nu=0}^{N-1}\frac{(-1)^\nu\Gamma(\mu+\nu+\frac{M-N+2-d}{2})(\psi(\mu+\nu+\frac{M-N+2-d}{2})-\ln2)}{2^{\mu}\Gamma(\nu+1)\Gamma(N-\nu)\Gamma(\nu+M-N+1)\Gamma(\mu+\frac{1-d}{2})}\\
\nonumber
\fl
&&+\frac{2^M\Gamma(\frac{M+1}{2})\Gamma(\frac{N+1}{2})}{M N\sqrt{\pi}}\sum_{\nu=0}^{N-1}\frac{(-2)^\nu \Gamma(\nu+\frac{M-N+3}{2})\psi(\nu+\frac{M-N+3}{2})}{\Gamma(N-\nu)\Gamma(\nu+M-N+1)\Gamma(\nu+1)}\\
\fl
&&+(d-1)\frac{2^M\Gamma(\frac{M+1}{2})\Gamma(\frac{N+1}{2})}{M N\sqrt{\pi}} \sum_{\nu=0}^{N-1}\frac{(-2)^{\nu}\,\Upsilon(\nu+\frac{M-N+1}{2})}{\Gamma(N-\nu)\Gamma(\nu+M-N+1)\Gamma(\nu+1)}.
\end{eqnarray}
Here, 
\begin{equation} 
 (m)_n=\frac{\Gamma(m+n)}{\Gamma(m)}
\end{equation}   
represents the Pochhammer function,
\begin{equation}
d=\mathrm{mod}(M-N,2),
\end{equation}
and
\begin{equation}
\Upsilon(k)=\left|\frac{\partial}{\partial \alpha} \left(\frac{\Gamma(2\alpha+2)}{2^{2\alpha}\Gamma(\alpha+2)}~_2F_1(\alpha+1,\alpha+\frac{3}{2},\alpha+2;-1)\right)\right|_{\alpha=k}.
\end{equation} 
 For $\beta=1$, odd $N$, the result is comparatively simple. We have
\begin{eqnarray}
\label{ent1o}
\nonumber
\fl
E^{(1)}(M,N)&=&\psi\Big(\frac{M N}{2}+1\Big)+\frac{(2M+2N-M N-2)}{M N}\psi(M)-\frac{(M+N-1)}{M N}\psi\Big(\frac{M}{2}\Big)\\
\nonumber
\fl
&&\!\!\!+\frac{(M-1)(N-1)}{M N}\ln(2)-\frac{(N^2-N+4)}{2M N}\\
\fl
&&\!\!\!-\frac{2^M\Gamma\left(\frac{M+1}{2}\right)\Gamma\left(\frac{N+1}{2}\right)}{M N \sqrt{\pi}}\sum_{\nu=0}^{N-1}\frac{(-2)^\nu\Gamma\left(\nu+\frac{M-N+3}{2}\right)\psi(\nu+M-N+1)}{\Gamma(N-\nu)\Gamma(\nu+1)\Gamma(\nu+M-N+1)}.~~~~
\end{eqnarray}
Finally, for $\beta=4$ we have
\begin{eqnarray}
\label{ent4}
\nonumber
\fl
E^{(4)}(M,N)&=&\sum _{\mu=2M+1}^{2M N} \frac{1}{\mu}-\frac{(2N-1)}{4M}+2^{2N-2}\Gamma(M)\Gamma(N)\\
\fl
&&\!\!\!\times\sum_{\mu=0}^{N-1}\sum_{\nu=\mu}^{2N}\frac{(-1)^{\nu} (\nu+2M-2N+1)(-\nu-1)_{2\mu}\psi(\nu+2M-2N+2)}{2^{2\mu}\Gamma(\mu+1) \Gamma(\nu+1)\Gamma(2N + 1-\nu)\Gamma(\mu+M-N+1)}.~~
\end{eqnarray}
The above expressions can be easily implemented in symbolic manipulation software packages like Mathematica.

For large $M, N$, (\ref{Ebeta_asym}) derives from (\ref{ent2}), (\ref{ent1e}), (\ref{ent1o}) and (\ref{ent4}) in the following way. In (\ref{ent2}) the summation term gives $\ln N$ thereby reproducing (\ref{Ebeta_asym}). In (\ref{ent1e}) the first four terms along with (\ref{psi_asym}) give back (\ref{Ebeta_asym}). In (\ref{ent1o}) the first five terms give (\ref{Ebeta_asym}) in the above limit. Similarly in (\ref{ent4}) the first two terms lead to (\ref{Ebeta_asym}). For $\beta=1$ and 4 the remaining summation terms become negligible in this limit.

The average von Neumann entropy for the three invariant cases is shown in figure \ref{avne}. Table I displays the numerical values for several $M, N$ values. It is clear that maximum average entanglement is obtained for $\beta=4$. 

%--------------------------------------------------------
\begin{table}%[h!b!p!]
\begin{center}
\renewcommand{\tabcolsep}{0.4cm}
\renewcommand{\arraystretch}{1.5}
  \begin{tabular}{ |c | c || c | c | c | }
    \hline
    $M$ & $N$ & $\beta$=1 & $\beta$=2 & $\beta$=4 \\ \hline\hline\hline
    
    2 & 2 & 0.27718 & 0.33333 & 0.37063 \\ \hline\hline\hline
    
    3 & 2 & 0.39772 & 0.45000 & 0.48099 \\ \hline
    3 & 3 & 0.59987 & 0.66230 & 0.69918 \\ \hline\hline\hline
    
    4 & 2 & 0.46451 & 0.50952 & 0.53490 \\ \hline
    4 & 3 & 0.71324 & 0.76988 & 0.80167 \\ \hline
    4 & 4 & 0.86452 & 0.92240 & 0.95494 \\ \hline\hline\hline
    
    5 & 2 & 0.50680 & 0.54563 & 0.56692 \\ \hline
    5 & 3 & 0.78503 & 0.83490 & 0.86210 \\ \hline
    5 & 4 & 0.96152 & 1.01441 & 1.04329 \\ \hline
    5 & 5 & 1.08065 & 1.13262 & 1.16108 \\ \hline\hline\hline
    
    6 & 2 & 0.53595 & 0.56988 & 0.58815 \\ \hline
    6 & 3 & 0.83442 & 0.87844 & 0.90204 \\ \hline
    6 & 4 & 1.02846 & 1.07596 & 1.10143 \\ \hline
    6 & 5 & 1.16366 & 1.21165 & 1.23744 \\ \hline
    6 & 6 & 1.26130 & 1.30789 & 1.33298 \\ \hline
  \end{tabular}
  \end{center} 
  \caption{Average von Neumann entropy evaluated to 5 decimal places for several $M\geq N$ values.}
  \label{Table2}
 \end{table}

If for $\beta=4$ we consider the Kramers degeneracy explicitly, the result for average von Neumann entropy will change to
\begin{equation}
E^{(4)*}(2M,2N)=E^{(4)}(M,N)+\ln2. 
\end{equation}
In this case the Hilbert spaces $\mathcal{H}_A$ and $\mathcal{H}_B$ are of dimensions $2N$ and $2M$ respectively with sum of all $2N$ eigenvalues being $1$. Again we find $E^{(4)*}(2M,2N)>E^{(2)}(2M,2N)>E^{(1)}(2M,2N)$ and $E^{(4)*}(2M,2N)/2>E^{(2)}(M,N)>E^{(1)}(M,N)$.

%~~~~~~~~~~~~~~~~~~~~~ Section 6 ~~~~~~~~~~~~~~~~~~~~~~~~

\section {Conclusion}
\label{secC}

We have given exact results for spectral density and average von Neumann entropy for the random pure states belonging to one of the three invariant classes of random matrix ensembles. We have shown using von Neumann entropy measure that, as far as the average is concerned, maximum entanglement is achieved if one generates states from the symplectic invariant class. Several ways of generating random pure states have been proposed by various authors \cite{BL2002,KTA2006,ZZF,Scot,GG,WH,Zni}. It will be of interest to test the theoretical predictions presented here using these and newer methods.

%~~~~~~~~~~~~~~~~~~~~~~~~~ Acknowledgements ~~~~~~~~~~~~~~~~~~~~~~~~~~~
\ack

The authors would like to thank Arul Lakshminarayan for fruitful discussions. SK acknowledges CSIR India for financial assistance.

% ~~~~~~~~~~~~~~~~~~~~~~~~~~~ Appendices ~~~~~~~~~~~~~~~~~~~~~~~~~~~~~~~

\appendix
\section{Relation between normalization constants in \texorpdfstring{(\ref{C})}{(4)} and \texorpdfstring{(\ref{Cft})}{(21)}}
\label{CCft}
\setcounter{section}{1}

First we find out the ratio between the normalizations in the jpd's given by (\ref{jpd}) and (\ref{ftjpd}). We have from (\ref{ftjpd}),
\begin{equation}
 (\mathcal{C}_{M,N}^{(\beta)})^{-1}=\int_0^\infty\cdots\int_0^\infty\delta\Big(\sum_{j=1}^N\lambda_j-1\Big) |\Delta_N(\{\lambda\})|^\beta \prod_{k=1}^N \lambda_k^\alpha.
\end{equation}
We introduce an auxiliary variable $t$ in the above expression and define $G(t)$ as,
\begin{equation}
G(t)=\int_0^\infty\cdots\int_0^\infty\delta\Big(\sum_{j=1}^N\lambda-t\Big) |\Delta_N(\{\lambda\})|^\beta\prod_{j=1}^N \lambda_j^\omega,
\end{equation}
so that $ (\mathcal{C}_{M,N}^{(\beta)})^{-1}=G(1)$. Taking the Laplace transform ($t\rightarrow s$) of $G(t)$ we get
\begin{equation}
\widetilde{G}(s)=\int_0^\infty\cdots\int_0^\infty |\Delta_N(\{\lambda\})|^\beta \prod_{j=1}^N \lambda_j^\omega e^{-s\lambda_j}. 
\end{equation}
We now substitute $s\lambda_j=x_j$ and obtain,
\begin{eqnarray}
\fl
\widetilde{G}(s)=s^{-\frac{\beta M N}{2}}\int_0^\infty\cdots\int_0^\infty |\Delta_N(\{x\})|^\beta \prod_{j=1}^N x_j^\omega e^{-x_j}=s^{-\frac{\beta M N}{2}}(\mathrm{C}_{M,N}^{(\beta)})^{-1},
\end{eqnarray}
using (\ref{Cft}).
Using the inverse Laplace transform result ($s\rightarrow t$)
\begin{equation}
\label{ilt1}
\mathcal{L}^{-1}[s^\alpha]=\frac{t^{-\alpha-1}}{\Gamma(-\alpha)},
\end{equation}
we obtain
\begin{equation}
 G(t)=\frac{t^{MN-1}}{\Gamma(\frac{\beta MN}{2})}(\mathrm{C}_{M,N}^{(\beta)})^{-1}.
\end{equation}
Finally, for $t=1$ we get
\begin{equation}
\label{ratio}
\frac{\mathcal{C}_{M,N}^{(\beta)}}{\mathrm{C}_{M,N}^{(\beta)}}=\Gamma\Big(\frac{\beta MN}{2}\Big).
\end{equation}
Equation (\ref{ratio}) may also be obtained using the auxialiary gamma function integral method discussed in \ref{AGFIM}.

% ~~~~~~~~~~~~~~~~~~~~~~~~~~~~~~~~~~~~~~~~~~~~~~~~~~~~~~~

\section{Relation between level densities in \texorpdfstring{(\ref{R1})}{(5)} and \texorpdfstring{(\ref{R1ft})}{(22)}}
\label{R1R1ft}
\renewcommand{\theequation}{B.\arabic{equation}}
\setcounter{equation}{0}

The level density for the Schmidt eigenvalues is (we drop the arguments $M, N$ in the following)
\begin{equation}
\fl
\mathcal{R}_1^{(\beta)}(\lambda_1)=N\int_0^\infty d\lambda_2\cdots\int_0^\infty d\lambda_N\mathcal{C}_N\delta\Big(\sum_{j=1}^N\lambda_j-1\Big) |\Delta_N(\{\lambda\})|^\beta\prod_{k=1}^N \lambda_k^\omega.
\end{equation}
We introduce the auxiliary variable $t$ and consider $\mathrm{R}_1^{(\beta)}(\lambda_1;\tau)$ defined by
\begin{equation}
\fl
\mathrm{R}_1^{(\beta)}(\lambda_1;t)=N\int_0^\infty d\lambda_2\cdots\int_0^\infty d\lambda_N\mathcal{C}_N\delta\Big(\sum_{j=1}^N\lambda_j-t\Big) |\Delta_N(\{\lambda\})|^\beta\prod_{k=1}^N \lambda_k^\omega.
\end{equation}
Thus, $\mathcal{R}_1^{(\beta)}(\lambda)=\mathrm{R}_1^{(\beta)}(\lambda;1)$.
Laplace transform ($t\rightarrow s$) of the above expression yields
\begin{equation}
\fl
\widetilde{\mathrm{R}}_1^{(\beta)}(\lambda_1;s)=N\int_0^\infty d\lambda_2\cdots\int_0^\infty d\lambda_N\,\mathcal{C}_N |\Delta_N(\{\lambda\})|\prod_{j=1}^N \lambda_j^\omega e^{-s \lambda_j}.
\end{equation}
This, on using (\ref{jpd}) and (\ref{R1}), gives
\begin{equation}
\label{rel}
\fl
\widetilde{\mathrm{R}}_1^{(\beta)}(\lambda;s)=\frac{\mathcal{C}_N}{C_N}\frac{1}{s^{\frac{\beta MN}{2}-1}}R_1^{(\beta)}(s\lambda)=\Gamma\Big(\frac{\beta MN}{2}\Big)s^{1-\frac{\beta MN}{2}} R_1^{(\beta)}(s\lambda),
\end{equation}
Thus the level density for fixed trace ensemble can be obtained from the level density of ordinary Wishart-Laguerre ensemble by taking inverse Laplace transform ($s\rightarrow t$) of the expression on right hand side of (\ref{rel}), and by substituting $t=1$.

% ~~~~~~~~~~~~~~~~~~~~~~~~~~~~~~~~~~~~~~~~~~~~~~~~~~~~~~~

\section{Proofs of equations \texorpdfstring{(\ref{R1ft2})$-$(\ref{R1ft4})}{(31)-(35)}}
\label{Proofs}
\renewcommand{\theequation}{C.\arabic{equation}}
\setcounter{equation}{0}

We consider the expression
\begin{equation}
\label{R1_ue_alt}
\fl
R_1^{(2)}(N;a;x)=w_{2a+1}(x) \frac{\Gamma(N+1)}{\Gamma(N+2a+1)}\big[L_{N-1}^{(2a+2)}(x)L_{N-1}^{(2a+1)}(x)-L_{N-2}^{(2a+2)}(x)L_{N}^{(2a+1)}(x)\big]
\end{equation}
for level density given in (\ref{R1_ue}). Note that $L_j^{(k)}(x)=0$ for $j<0$. The above expression (and other equivalent ones) can be obtained from (\ref{R1_ue}) using the results given in~\cite{Sz}. We now use the following standard expansion~\cite{Sz} of Laguerre polynomial in (\ref{R1_ue_alt}):
\begin{equation}
\label{Laguerre}
L_n^{(b)}(x)=\sum_{\mu=0}^n \frac{\Gamma(n+b+1)(-x)^\mu}{\Gamma(b+\mu+1)\Gamma(n-\mu+1)\Gamma(\mu+1)} .
\end{equation}
This gives an expression for the level density (\ref{R1_ue}) involving double sum. Substituting this expression in (\ref{rel}) and performing the inverse Laplace transform using the result
\begin{equation}
\label{ilt2}
\mathcal{L}^{-1}[s^\gamma e^{s \lambda}]=\frac{(t-\lambda)^{-\gamma-1}\Theta(t-\lambda)}{\Gamma(-\gamma)},
\end{equation}
gives the result for level density of Schmidt eigenvalues. This result is a double sum, however one of the sums can be performed in terms of hypergeometric function defined in (\ref{Smu}), thus leaving the final result (\ref{R1ft2}) as a single sum.

To derive (\ref{R1ft1e}), (\ref{R1ft1o}) we use again (\ref{R1_ue_alt}) and the expansion (\ref{Laguerre}) for Laguerre polynomial in (\ref{R1_oe_e}) and (\ref{R1_oe_o}). Along with (\ref{ilt2}) we also need the following Laplace inverse result:
\begin{equation}
\label{ilt3}
\fl
\mathcal{L}^{-1}[s^\gamma e^{s \lambda} \Gamma(\alpha+1,s \lambda)]=(t-2\lambda)^{-\alpha-\gamma-1}\lambda^{\alpha}~_2\widetilde{F}_1\left(1,-\alpha;-\alpha-\gamma;2-\frac{t}{\lambda}\right)\Theta(t-2\lambda). 
\end{equation}

Equation (\ref{R1ft4}) can similarly be derived from (\ref{R1_se}).

% ~~~~~~~~~~~~~~~~~~~~~~~~~~~~~~~~~~~~~~~~~~~~~~~~~~~~~~~

\section{Auxiliary gamma function intergral method}
\label{AGFIM}
\renewcommand{\theequation}{D.\arabic{equation}}
\setcounter{equation}{0}

We introduce an auxiliary gamma function intergral in (\ref{E1beta}) as
\begin{eqnarray}
\fl
\nonumber
E^{(\beta)}(M,N)&=&-\frac{\mathcal{C}_N}{\Gamma(\Omega)}\int_0^\infty e^{-r}r^{\Omega-1}\int_0^{\infty }\cdots\int_0^{\infty}\left(\sum_{k=1}^N\lambda_k \ln \lambda_k\right)\\
\fl
&&\times\delta\left(\sum_{i=1}^N\lambda_i-1\right)|\Delta_N(\{\lambda\})|^{\beta}\prod_{j=1}^{N} \lambda_j^\omega \,d\lambda_1...d\lambda_N\,dr.~~~~~~~~~~~~
\end{eqnarray}
This can be reduced to the following after some manipulations:
\begin{eqnarray}
\nonumber
\fl
E^{(\beta)}(M,N)=\psi(B)
-\frac{\mathcal{C}_N}{\Gamma(\Omega)}\int_0^\infty e^{-r}r^{\Omega-1}\int_0^{\infty }\cdots\int_0^{\infty}\left(\sum_{k=1}^N\lambda_k \ln (r\lambda_k)\right)\delta\left(\sum_{i=1}^N\lambda_i-1\right)\\
\fl
\hspace{3cm} \times|\Delta_N(\{\lambda\})|^{\beta}\prod_{j=1}^{N} \lambda_j^\omega \,d\lambda_1...d\lambda_N\,dr.~~~~~~~~~~~~
\end{eqnarray}
where $\psi(x)$ is the digamma function as defined in (\ref{digamma}). Now substituting $\lambda_j=x_j/r$, we obtain
\begin{eqnarray}
\nonumber
\fl
E^{(\beta)}(M,N)\!\!\!\!&\!\!\!\!=&
 \psi(\Omega)-\frac{\mathcal{C}_N}{\Gamma(\Omega)}\int_0^\infty e^{-r}r^{\Omega-1}\int_0^{\infty }\cdots\int_0^{\infty}\left(\sum_{k=1}^N\frac{x_k}{r} \ln x_k\right)\delta\left(\sum_{i=1}^N\frac{x_i}{r}-1\right)\\
\nonumber
\fl
 &&\times\left|\Delta_N\left(\Big\{\frac{x_1}{r}\Big\}\right)\right|^{\beta}\prod_{j=1}^{N} \left(\frac{x_j}{r}\right)^\omega \,\frac{dx_1}{r}...\frac{dx_N}{r}\,dr\\
\nonumber
\fl
&=&\psi(\Omega)-\frac{\mathcal{C}_N}{\Gamma(\Omega)}\int_0^\infty \frac{e^{-r}r^{\Omega-1}}{r^{\frac{\beta M N}{2}}}\int_0^{\infty }\cdots\int_0^{\infty}\left(\sum_{k=1}^N x_k \ln x_k\right)\delta\left(\sum_{i=1}^N x_i-r\right)\\
&& \times|\Delta_N(\{x\})|^{\beta}\prod_{j=1}^{N} x_j^\omega \, dx_1...dx_N\,dr.
\end{eqnarray}
The choice $\Omega=\beta MN/2+1$ gives
\begin{eqnarray}
\nonumber
\fl
E^{(\beta)}(M,N)=\psi\left(\frac{\beta M N}{2}+1\right)-\frac{\mathcal{C}_N}{\Gamma(\frac{\beta M N}{2}+1)}\int_0^\infty e^{-r}\int_0^{\infty }\cdots\int_0^{\infty}\left(\sum_{k=1}^N x_k \ln x_k\right)\\
\nonumber
 \times\delta\left(\sum_{i=1}^N x_i-r\right)|\Delta_N(\{x\})|^{\beta}\prod_{j=1}^{N} x_j^\omega \, dx_1...dx_N\,dr~~~~~~~~~~~~
\end{eqnarray}
\begin{eqnarray}
\nonumber
\fl
=\psi\left(\frac{\beta M N}{2}+1\right)-\frac{\mathcal{C}_N}{\Gamma(\frac{\beta M N}{2}+1)}\int_0^{\infty }\cdots\int_0^{\infty}\left(\sum_{k=1}^N x_k \ln x_k\right)\\
\nonumber
 \times|\Delta_N(\{x\})|^{\beta}\prod_{j=1}^{N} x_j^\omega e^{-x_j} \, dx_1...dx_N~~~~~~~~~~~~
\end{eqnarray}
\begin{eqnarray}
\nonumber
\fl
=\psi\left(\frac{\beta M N}{2}+1\right)
-\frac{1}{\Gamma(\frac{\beta M N}{2}+1)}\frac{\mathcal{C}_N}{C_N}\int_0^{\infty }\cdots\int_0^{\infty}\left(\sum_{k=1}^N x_k \ln x_k\right)P(\{x\}) \, dx_1...dx_N.\\
\end{eqnarray}
Using the ratio of constants, $\mathcal{C}_N/C_N$, from (\ref{ratio}) we eventually obtain (\ref{E2beta}).

% ~~~~~~~~~~~~~~~~~~~~~~~~~~~ References ~~~~~~~~~~~~~~~~~~~~~~~~~~~~~~~

\end{document}